\documentclass[11pt]{article}

\usepackage[utf8]{inputenc}
\usepackage[T1]{fontenc}
\usepackage[a4paper,margin=1in]{geometry}
\usepackage{graphicx}
\usepackage{amsmath}
\usepackage{amssymb}
\usepackage{textcomp}
\usepackage{booktabs}
\usepackage[font=small,labelfont=bf]{caption}
\usepackage[authoryear,round]{natbib}
\usepackage{aas_macros}
\usepackage[colorlinks=true,allcolors=blue]{hyperref}
\usepackage{journalnames}

\graphicspath{{figures/}}

\newcommand{\degr}{\ensuremath{^{\circ}}}
\newcommand{\chem}[1]{\ensuremath{\mathrm{#1}}}

\newcommand{\msinv}{\ensuremath{\,\mathrm{m\,s^{-1}}}}
\newcommand{\um}{\ensuremath{\,\mu\mathrm{m}}}

\title{\textbf{Giant Planet Atmospheres}}
\author{Henrik Melin \\Northumbria University, Newcastle upon Tyne, UK. \\
\ \texttt{henrik.melin@northumbria.ac.uk}}
\date{}

\begin{document}

\maketitle

\begin{center}
\footnotesize
This is the author's version of an article accepted for publication in the
\textit{Oxford Research Encyclopedia of Planetary Science} (Oxford University
Press).\\
DOI: \href{https://doi.org/10.1093/9780197851500.003.0345}{10.1093/9780197851500.003.0345}
\end{center}

\begin{abstract}
\noindent
The giant planets, Jupiter, Saturn, Uranus, and Neptune, all have vibrant and
dynamic atmospheres. The iconic belt--zone structure of Jupiter, together with
the Great Red Spot, is instantly recognizable. Saturn, with its dramatic ring
system and more muted atmosphere, is a formidable jewel in the Solar System. In
the outer reaches, the pale blue Uranus and Neptune are found, worlds about
which ultimately very little is known. The atmospheres of these planets are
dominated by hydrogen and helium, and unlike the Earth, they do not have a solid
surface. These differences generate inherently different types of atmospheres,
but there are also similarities. For example, the condensation of water, which
forms the familiar clouds on Earth, also occurs on the giant planets. Broadly
speaking, the atmosphere can be divided into different regimes defined by their
temperature gradients. In the troposphere, where weather occurs, the
temperatures decrease as a function of increasing altitude as convection moves
internal heat upward; the rising material expands and cools. Above this region
lies the stratosphere, defined by a positive temperature gradient, where
hydrocarbons are heated by ultraviolet radiation from the Sun (analogous to
ozone heating in the terrestrial stratosphere), which also drives substantial
photochemistry. This is followed by a mesosphere that cools as a function of
altitude, a region that is ill-defined at the giant planets. Finally, the upper
atmosphere connects to the space environment and is heated by both solar extreme
ultraviolet light and auroral processes. The giant planets are energized both by
internal heat and by solar heating. These energy inputs, along with the fast
rotation rates of these planets, drive dynamics by establishing global
circulation patterns and generating both waves and instabilities. They also
drive chemistry within the atmosphere, which alters composition and can influence
the temperature structure, particularly in the stratosphere. The atmospheres of
these planets can be observed using telescopes on the ground and in space, which
can reveal composition, dynamics, and long-term changes. Much of the detailed
knowledge of these planets comes from spacecraft such as the iconic Grand Tour
of Voyager 1 and 2 in the 1970s and 1980s, as well as the Galileo and Juno
missions to Jupiter and the Cassini mission to Saturn. In the 2030s, both the
European Space Agency's Jupiter Icy Moons Explorer (JUICE) mission and the NASA
Europa Clipper mission will arrive at Jupiter, and momentum within the planetary
science community is building toward a mission to Uranus that would arrive at
some point in the 2040s.

\end{abstract}

\section{Introduction}
\label{sec:intro}

\begin{figure}[ht]
\centering
\includegraphics[width=\linewidth]{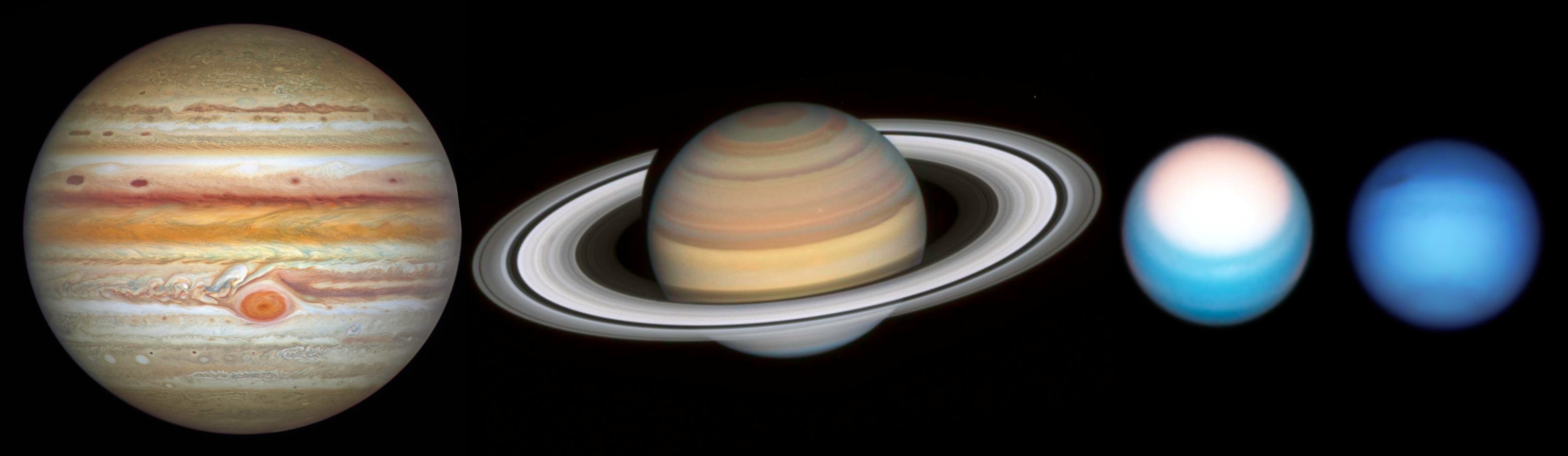}
\caption{A giant planet family portrait: Jupiter, Saturn, Uranus, and Neptune,
obtained with the Hubble Space Telescope in 2021. \textit{Source}: NASA, ESA,
A.~Simon (Goddard Space Flight Center), and M.~H.~Wong (University of California,
Berkeley) and the OPAL team.}
\label{fig:family}
\end{figure}

An atmosphere is a gaseous envelope confined by the gravitational forces of a
body. From the very thin and unstable atmosphere of Mercury to the dense and
vigorously churning atmosphere that envelops Jupiter, there is an extraordinary
variety of atmospheres observed throughout the Solar System. The nature of the
atmospheres of the planets (and moons) in the Solar System is shaped by
processes such as solar forcing, internal heating, and rotation rates, along
with the way these planets have evolved since their formation some 4.6 billion
years ago. The giant planets, Jupiter, Saturn, Uranus, and Neptune
(Figure~\ref{fig:family}), differ from the terrestrial planets in that they do
not have a distinct surface, have fast rotation rates, and are much larger and
heavier. The sheer size of these planets means that the strong gravitational
field can retain lighter species, and so the outer layers are dominated by
hydrogen and helium.

The giant planets orbit far from the Sun and therefore receive little sunlight:
Jupiter receives about 4\% and Neptune 0.1\% of the sunlight received on Earth.
Despite this, the energy provided by the Sun plays an important role, especially
in the stratosphere and the thermosphere. Here, photochemistry can drive heating
and complex chemistry, which can exhibit strong seasonal behavior. The tilt of
Jupiter's spin axis is small, only 3\degr, which means it has very muted seasonal
behavior, whereas the tilts of Saturn (27\degr), Uranus (97\degr), and Neptune
(28\degr) are larger, making seasonal geometry important for the thermal
structure and dynamics of the stratosphere and upper atmosphere, just as it is
for Earth's atmosphere.

The giant planets get their name by virtue of being very large, as summarized in
Table~\ref{tab:planets}, and they contain 99.6\% of the total planetary mass. By
virtue of their size and fast rotation rates, they also dominate the bulk angular
momentum within the Solar System, accounting for $\sim$90\% of it, with Jupiter
alone contributing $\sim$60\%. Ultimately, their size, distance from the Sun, and
composition provide constraints on the formation of the Solar System, and by
understanding the atmospheres of these planets, they can be placed in the context
of planetary evolution. The fact that observations of these atmospheres occur at
a particular point on their evolutionary journey must also be considered. The
distance at which a planet forms from its host star, the amount of mass that
accretes in the protoplanetary nebula, along with its evolution as a function of
distance from the star, are all important factors in determining what type of
planet it will become millions or billions of years down the line.

For a protoplanetary nebula to retain any chemical elements, the temperature must
be low enough for these species to condense into a solid state. Consequently, a
planet that forms close to its parent star will experience high temperatures, and
so volatile gases, such as water (\chem{H_2O}) and methane (\chem{CH_4}), are
unable to condense and will escape the system. This means that only elements with
higher melting points, such as iron, sodium, and silicates, can exist as solids.
This explains why the terrestrial planets are rich in heavy elements but lacking
lighter ones. Conversely, the farther a planet forms from its parent star, the
more condensable species it can retain. The distance beyond which volatile icy
compounds remain solid during the formation of planets is known as the
\textit{frost line}, which sits at $\sim$5~AU \citep{2007prpl.conf..863J}.

The purpose of this article is to provide the reader with an overview of the
atmospheres of the giant planets, the processes that occur within them, and the
general lexicon used to describe the relevant physics. Detailed mathematical
descriptions are omitted here, and the reader is referred to more comprehensive
works \citep{2009gpss.book.....I,sanchez2011introduction}. Section~\ref{sec:structure}
describes the basic vertical structure of these atmospheres, followed by the
processes that can alter the atmosphere over long timescales in
Section~\ref{sec:evolution}. Section~\ref{sec:dynamics} describes the basic
dynamics of these atmospheres, including convection and turbulence triggered by
instabilities. In Section~\ref{sec:gasgiants}, the gas giants (Jupiter and
Saturn) will be discussed, followed by the ice giants (Uranus and Neptune) in
Section~\ref{sec:icegiants}. Finally, Section~\ref{sec:conclusion} contains a
brief conclusion.

\begin{table}[ht]
\centering
\caption{Basic Physical Characteristics of the Giant Planets}
\label{tab:planets}
\begin{tabular}{lcccccc}
\toprule
Planet & Distance & Radius & Mass & Density & Orbital Period & Rotation\\
 & (AU) & ($R_\oplus$) & ($M_\oplus$) & (g\,cm$^{-3}$) & (years) & (hours)\\
\midrule
Jupiter & 5.2  & 11.0 & 318 & 1.3 & 11.9  & 9.9\\
Saturn  & 9.5  & 9.1  & 95  & 0.7 & 29.5  & 10.7\\
Uranus  & 19.2 & 4.0  & 14  & 1.3 & 84.1  & 17.2\\
Neptune & 30.0 & 3.9  & 17  & 1.6 & 164.8 & 16.1\\
\bottomrule
\end{tabular}
\end{table}

\section{Structure of an Atmosphere}
\label{sec:structure}

\begin{figure}[ht]
\centering
\includegraphics[width=0.8\linewidth]{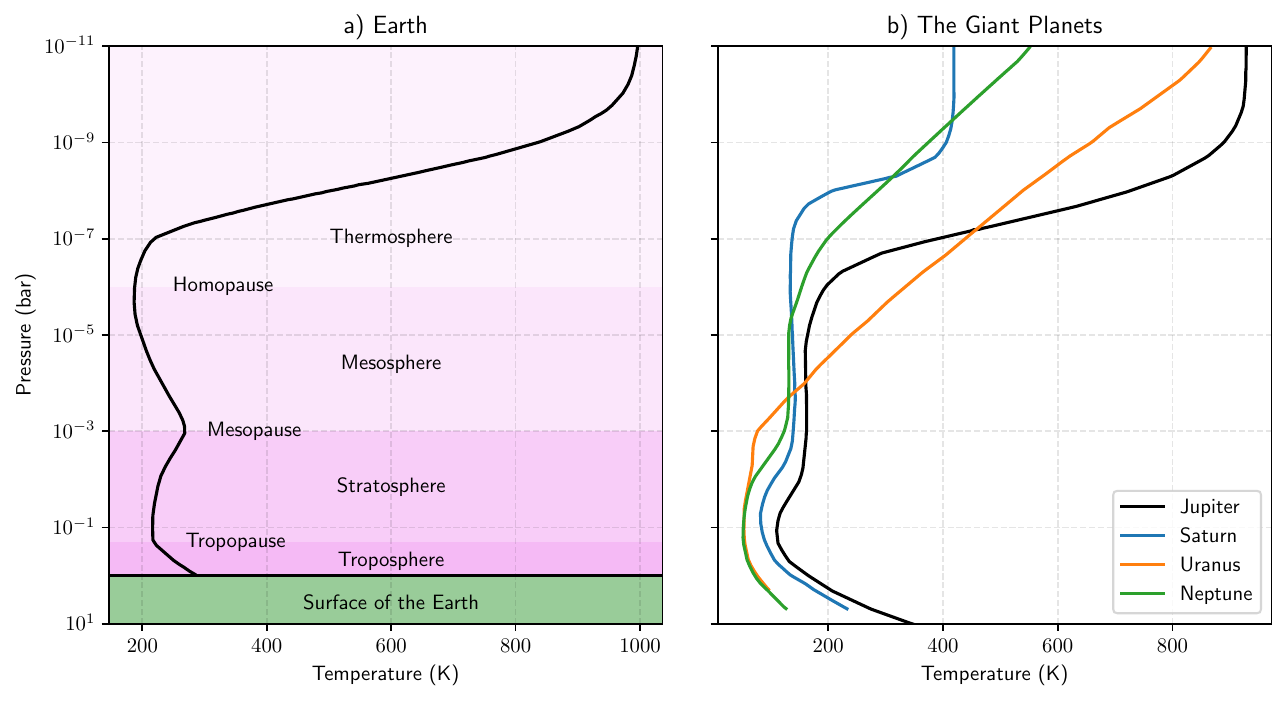}
\caption{(a) The vertical temperature profile of Earth. (b) The temperature
profiles of the giant planets. \textit{Source}: Adapted from
\citet{2022RemS...14.6326O}.}
\label{fig:tprofiles}
\end{figure}

An atmosphere can be divided into separate layers, delineated by distinct
boundaries largely defined by the behavior of the vertical temperature profile,
which in turn is driven by the different physical, dynamical, and chemical
processes within each region. These regions are categorized in the same way on
the giant planets as they are on Earth, as illustrated in Figure~\ref{fig:tprofiles}.
This section summarizes these regions, moving upward in altitude toward decreasing
pressures.

\subsection{Deep Interior}

On Earth, the solid surface absorbs sunlight, which in turn heats the atmosphere
at ground level. The giant planets do not have solid surfaces; however, there is
still a source of heat available in the deep atmosphere. This is energy left from
the formation of these bodies, and the continuing slow collapse of these planets
constantly releases heat, which warms their atmospheres at depth, through a
process known as the \textit{Kelvin--Helmholtz mechanism}. Jupiter, Saturn, and
Neptune all radiate strong internal heat, whereas Uranus has very little to no
internal heat \citep{1990Icar...84...12P}.

\begin{figure}[ht]
\centering
\includegraphics[width=0.8\linewidth]{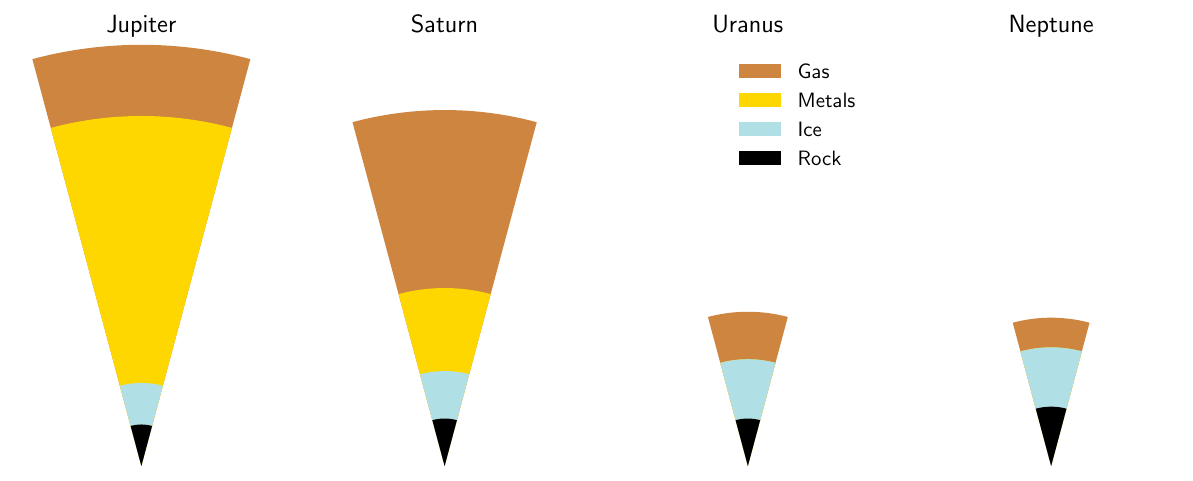}
\caption{The basic interior composition of the giant planets.}
\label{fig:interior}
\end{figure}

The basic interior structure of the giant planets is shown in
Figure~\ref{fig:interior}. The cores are thought to be made up of dense rocky
materials, followed by an envelope of ices, metallic hydrogen and helium in a
highly compressed environment, followed by the gaseous atmospheric envelope,
containing mostly hydrogen and helium. The convective metallic envelopes of the
cores are believed to be the origin of the powerful magnetic fields of these
planets, and if the convection changes over time, the observed magnetic fields
above the surface may also change on convective timescales. Therefore, these
types of observations of secular change in the magnetic field can reveal
long-term change deep inside the planet. Another way to sample the deep interior
of a planet is to observe how the gravitational force exerted on a spacecraft
flying close to the planet changes with position, revealing the \textit{gravity
moments} of the planet and, therefore, the internal structure. On Jupiter, such
measurements by the Juno spacecraft have revealed that it has an extended region
enriched in heavy elements \citep{2017GeoRL..44.4649W}.

\subsection{Troposphere}

The visual appearance of the giant planets is defined by sunlight reflected from
the clouds and hazes in the troposphere, the lower part of the atmosphere. This
is the weather layer, which is largely opaque and dominated by convection and
turbulence. Internal heat provides a source of heat at low altitudes, and as
parcels of air rise toward higher altitudes and lower pressures, the atmosphere
cools, producing a temperature profile that decreases sharply with altitude. At
the upper boundary of the troposphere lies the tropopause, the coldest part of a
giant planet atmosphere.

In the troposphere, as species rise in altitude toward cooler temperatures,
certain species will condense into solids and ices to form clouds, in a similar
way to how water vapor rises in Earth's troposphere and condenses to form clouds
at higher altitudes. Convection in the presence of these condensable species is
called \textit{moist convection} (as opposed to \textit{dry convection}). In the
molecular nitrogen- and oxygen-dominated atmosphere of Earth, the lighter water
vapor is naturally buoyant, whereas in the hydrogen- and helium-dominated
atmosphere of the giant planets, the condensable species tend to sink, and
therefore the formation of clouds becomes dependent on convection moving these
molecules toward higher, cooler altitudes at which they can condense. At the
giant planets, condensable species include water, ammonia, methane, as well as
more complex molecules such as ammonium hydrosulfide (\chem{NH_4SH}). The
formation of clouds releases heat, which modifies the vertical temperature
profile, because clouds can absorb and scatter radiation.

In addition to clouds, higher-altitude hazes and aerosols can also form in the
upper troposphere and lower stratosphere. \textit{Hazes} are evenly dispersed and
relatively uniform layers of small particles ($<$1\um\ in size), whereas
\textit{aerosols} are more complex hydrocarbon compounds, products of methane
photochemistry, that coagulate and condense. Aerosols can absorb significant
amounts of solar energy \citep{2015NatCo...610231Z}.

Uranus and Neptune, and to some extent Saturn, are shrouded in hazes in the lower
stratosphere that obscure the planet, hiding the intricate weather occurring
below. Strong convective events can lift clouds through this haze, making them
visible. Neptune is strongly convective and has bright storms that appear and
disappear on timescales of days and appear to be associated with the solar cycle
\citep{2023Icar..40415667C}. On Saturn, during northern spring, a massive storm
can develop, like a hurricane on Earth. The latest of these events occurred in
2010--2011 and was tracked by the Cassini spacecraft in orbit.

As the atmosphere becomes less dense with altitude, radiation becomes more
efficient at transporting energy, as opposed to convection, and the boundary
between these regimes is called the \textit{radiative--convective boundary},
which occurs in the upper troposphere.

\subsection{Stratosphere}

Above the temperature inversion at the tropopause lies the stratosphere, in which
the temperature increases as a function of altitude. This is driven by the
absorption of solar extreme ultraviolet photons by hydrocarbons, predominantly
methane, which can set off chains of complex photochemistry
\citep{2000Icar..143..244M}, generating species such as acetylene
(\chem{C_2H_2}) and ethane (\chem{C_2H_6}), which in turn provide a source of
radiative cooling. A similar temperature structure is seen on Earth, where
absorption of sunlight by ozone (\chem{O_3}) drives the heating. In other words,
the temperature profile in the radiative domain of the stratosphere is governed
largely by the solar-driven chemistry that takes place within it.

Because temperatures rise with altitude, vertical transport is strongly
inhibited, and therefore the movements within it tend to be very slow. In the
lower stratosphere, eddy, or turbulent, diffusion dominates, whereas in the upper
stratosphere, where the mean free path of the atmosphere becomes larger,
molecular diffusion becomes more important \citep{2007jupi.book..129M}.

At high latitudes, auroral processes provide an additional source of energy that
can produce heating and chemical reactions in the stratosphere, which can
penetrate deep into the polar atmosphere \citep{2017Icar..292..182S}.

\subsection{Mesosphere}

On Earth, as the ozone concentration decreases with altitude and carbon dioxide
(\chem{CO_2}) radiative cooling becomes more important, the temperature also
decreases with altitude. This defines the mesosphere, containing the coldest
temperatures seen in the telluric atmosphere. This decrease is not seen at the
giant planets, or if it exists, the mesosphere is extremely limited. This is
because the photochemical products of methane are distributed across a broad
range of altitudes, reaching the homopause.

\subsection{Upper Atmosphere}

The upper atmosphere has two components: the neutral thermosphere and the charged
particle ionosphere, produced by the ionization of the atmospheric constituents
either by solar extreme ultraviolet (EUV) radiation or by charged particles
delivered along magnetic field lines as part of the auroral process
\citep[e.g., see][]{MagnetosphereIonosphereCoupling,PlanetaryAurorae}. The lower
boundary of the upper atmosphere is the homopause, at which eddy diffusion equals
molecular diffusion. In other words, this is the boundary above which turbulent
mixing is no longer dominant, and the species fall out according to their scale
heights, which in turn are governed by their molecular weights. This means that
the heavier elements (i.e., hydrocarbons) are distributed very close to the
homopause, whereas lighter species, such as atomic and molecular hydrogen, can
extend for thousands of kilometers at very low densities.

The upper atmospheres of the giant planets are all observed to be very hot, and
it is unknown why. For Earth, the temperature of the upper atmosphere can be
accounted for by solar EUV radiation. In contrast, the upper atmospheres of the
giant planets are hundreds of Kelvins hotter than solar input alone can account
for, and this discrepancy between observations and models has been dubbed the
\textit{energy crisis}. Suggested solutions are the global redistribution of
auroral energy \citep{2021Natur.596...54O} and the breaking of gravity waves
\citep{2016Natur.536..190O}, but to date, no clear consensus exists.

\subsection{Exosphere}

The exosphere, delineated from the upper atmosphere by the exobase, marks the
boundary between the atmosphere and the space environment. Here, the density is
extremely low, and at the exobase the mean free path of a molecule is equal to
that of the scale height.

\section{Atmospheric Evolution}
\label{sec:evolution}

Once a planet is formed out of the protoplanetary disk, there are a number of
processes that can change the composition of its atmosphere over long timescales:

\begin{itemize}
\item \textbf{Jeans' escape:} The kinetic energy of a molecule can be increased
by several processes within an atmosphere, the strongest of which are absorption
of solar radiation and dissipation of gravity waves. A molecule with sufficiently
large velocity can overcome the gravitational forces that confine it and escape
the atmosphere. This process is clearly more efficient for lighter molecules.
This escape can, over time, significantly alter the composition of terrestrial
planets and moons, but since the giant planets are so massive, this process is
unlikely to have significantly shaped the evolution of these systems since their
formation.

\item \textbf{Impacts:} For Earth, the bombardment by icy planetesimals delivered
large amounts of water early in the history of the Solar System that radically
changed the atmospheric composition and ultimately enabled the germination of
life. This highlights how impacts can fundamentally alter atmospheric
composition. The deep gravitational wells of the giant planets mean that they can
easily attract small interplanetary bodies, such as comets and asteroids. The
most famous example of this is the Shoemaker--Levy 9 impact on Jupiter in July
1994, where the large, fragmented comet (the largest fragment being 1.8~km in
diameter) slammed into the southern hemisphere of Jupiter. These types of events
can deliver new materials, eject materials already present, as well as drive
localized chemistry.

\item \textbf{Stellar Luminosity:} Both the evolution of a star and the process
of planetary migration, where the orbital distance of a planet changes, can
modify the amount of stellar flux that a planet receives. This in turn provides a
variable energy input into a planet's atmosphere, which can drive changes in
Jeans' escape and increase or decrease photochemistry, which can, for example,
alter the stratospheric temperature profile.

\item \textbf{Ionospheric Escape:} On Earth, the charged ionosphere is
magnetically connected to plasma in the magnetic field. Since the pressure is
high in the ionosphere compared to the magnetic tail downstream of the planet,
there is a tendency for plasma to move from the ionosphere to the tail, known as
the \textit{polar wind}. A similar process is thought to occur at the giant
planets, and modeling has suggested that $\sim$1.5$\times$10$^{28}$ protons per
second could be lost at Jupiter \citep{2020JGRA..12527727M}, which is a similar
number of charged particles sourced from the volcanic plumes of Io
\citep{1997GeoRL..24.2111B}.
\end{itemize}

\section{Atmospheric Dynamics}
\label{sec:dynamics}

The motion of an atmosphere is highly complex and can sometimes appear to be
chaotic and random. This section will discuss the broadscale movements of the
atmospheres of the giant planets and the general principles that can be applied.

\subsection{Description of the Mathematical Toolkit}

Consider a self-contained parcel of air in which energy does not leave or enter
(i.e., it is adiabatic) and that has a constant entropy (i.e., it is isentropic).
The foundations of circulation within an atmosphere can then be explored using
the \textit{ideal gas law}:
\begin{equation}
PV = nRT
\end{equation}
where $P$ is the pressure, $V$ is the volume, $n$ is the number density, $R$ is
the ideal gas constant ($R = 8.314$~J\,K$^{-1}$\,mol$^{-1}$), and $T$ is the
temperature. The fundamental driver of atmospheric circulation at the giant
planets is the heating of the atmosphere at depth. A parcel of air deep in the
atmosphere that is being heated will grow in volume and thus decrease in density.
This change will drive the parcel to move upward in altitude toward lower
pressures. As it moves upward, the decreasing pressures will act to further
increase the volume, cooling it. Once the parcel has equilibrated with its
surroundings in the upper troposphere, the reverse process can take place: a cold
parcel contracts and moves toward higher pressures, completing the convection
pattern of upwelling and downwelling.

The rate at which the temperature of an adiabatic parcel of air changes as a
function of altitude is described by either the \textit{dry} or \textit{moist
adiabatic lapse rate}, which refers to the absence or presence of condensable
species within the parcel. These expressions provide a tool with which to
calculate idealized vertical temperature profiles.

There are a number of equations that govern the motion of an atmosphere. The
principal of these is the \textit{Navier--Stokes equation}, which balances the
forces exerted on a parcel of air, namely \textit{gravity, friction, the pressure
gradient}, and \textit{centrifugal forces}. The pressure gradient is a result of
differing pressures between different or within parcels of air, with the flow of
air moving from low pressure toward higher pressures. The centrifugal forces give
rise to \textit{Coriolis forces}, which restrict movement of the atmosphere in
the meridional (north--south) direction. Due to the size and fast rotation of the
giant planets, the Coriolis force places strict limits on the meridional mixing
of the atmosphere.

The \textit{geostrophic equations} describe the motion of the atmosphere, which
is dependent on the pressure gradient in each direction. This description
requires the Coriolis force to be significantly larger than the pressure gradient
force, and satisfying this condition is called the \textit{geostrophic
approximation}. The condition can be parameterized into a single number called
\textit{the Rossby number}, where values much less than one indicate that the
Coriolis forces dominate. In these conditions, winds flow along lines of constant
pressure, rather than from low to high pressures, as would be expected due to
pressure gradient forces. The rapid rotation of these planets suggests that this
approximation holds.

The \textit{thermal wind equations} relate the vertical shear of the geostrophic
wind to the vertical temperature gradients. If the zonal velocity is known at a
particular altitude (see Section~\ref{sec:zonalwinds}), then this relationship can
be used to derive the vertical wind field based on the observed vertical
temperature gradients.

There are two types of vortices in an atmosphere: \textit{cyclones}, which form
over a region of low pressure, and \textit{anticyclones}, which form over a region
of high pressure. They have opposite senses of rotation, and the direction of
motion is dependent on the hemisphere in which they appear. The most famous
anticyclone is the Great Red Spot (GRS) on Jupiter. These can be characterized in
terms of \textit{vorticity}, which is a measure of rotational speed and is related
to angular momentum, which is a conserved quantity. \textit{Potential vorticity}
is a related concept that is also conserved and that can predict the behavior of
vortical flows as they evolve (e.g., change in size).

\begin{figure}[ht]
\centering
\includegraphics[width=0.8\linewidth]{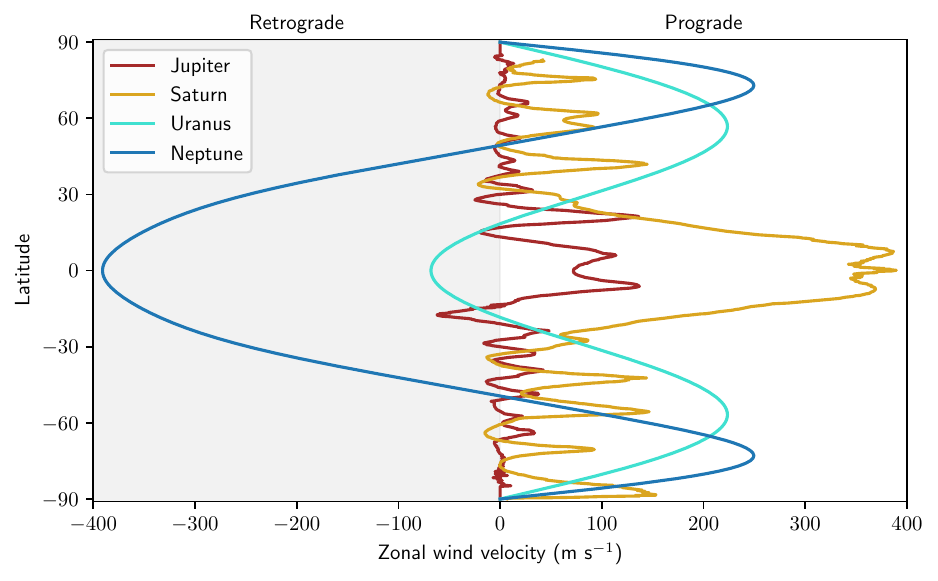}
\caption{The observed zonal wind velocities on Jupiter
\citep{2003Sci...299.1541P}, Saturn \citep{2011Icar..215...62G}, Uranus
\citep{2001Icar..153..229H}, and Neptune \citep{1998Icar..136...27F}.}
\label{fig:zonalwinds}
\end{figure}

\subsection{Zonal Winds}
\label{sec:zonalwinds}

The observed zonal (i.e., east--west) winds in the troposphere for the four giant
planets are shown in Figure~\ref{fig:zonalwinds}. These are derived from cloud
tracking, observing the relative motion of clouds or other features between
successive rotations. Observations have shown that these winds are very stable
over time, and how these extremely strong winds can be sustained over very long
timescales is not known. Their stability indicates that they are driven by
internal processes rather than external solar forcing. The equators of Jupiter
and Saturn have prograde (superrotating) zonal winds, whereas the equators of
Uranus and Neptune have retrograde (subrotating) zonal winds. The strongest winds
are seen on Neptune, followed by Saturn and then Uranus, all of which have faster
winds than Jupiter.

On Jupiter, the zonal wind structure is present at great depth ($>$10$^5$~bar;
\citealp{2023NatAs...7.1463K}), suggesting that the observed winds may be a
signature of processes that occur deep inside the planet. By using the thermal
wind equation, the way the winds decay with height can be calculated, and on all
four giant planets they become zero at $\sim$4 scale heights above the cloud
tops. At Jupiter and Saturn, the peak velocities in the zonal flows coincide with
the boundaries between belts and zones.

\begin{figure}[ht]
\centering
\includegraphics[width=0.8\linewidth]{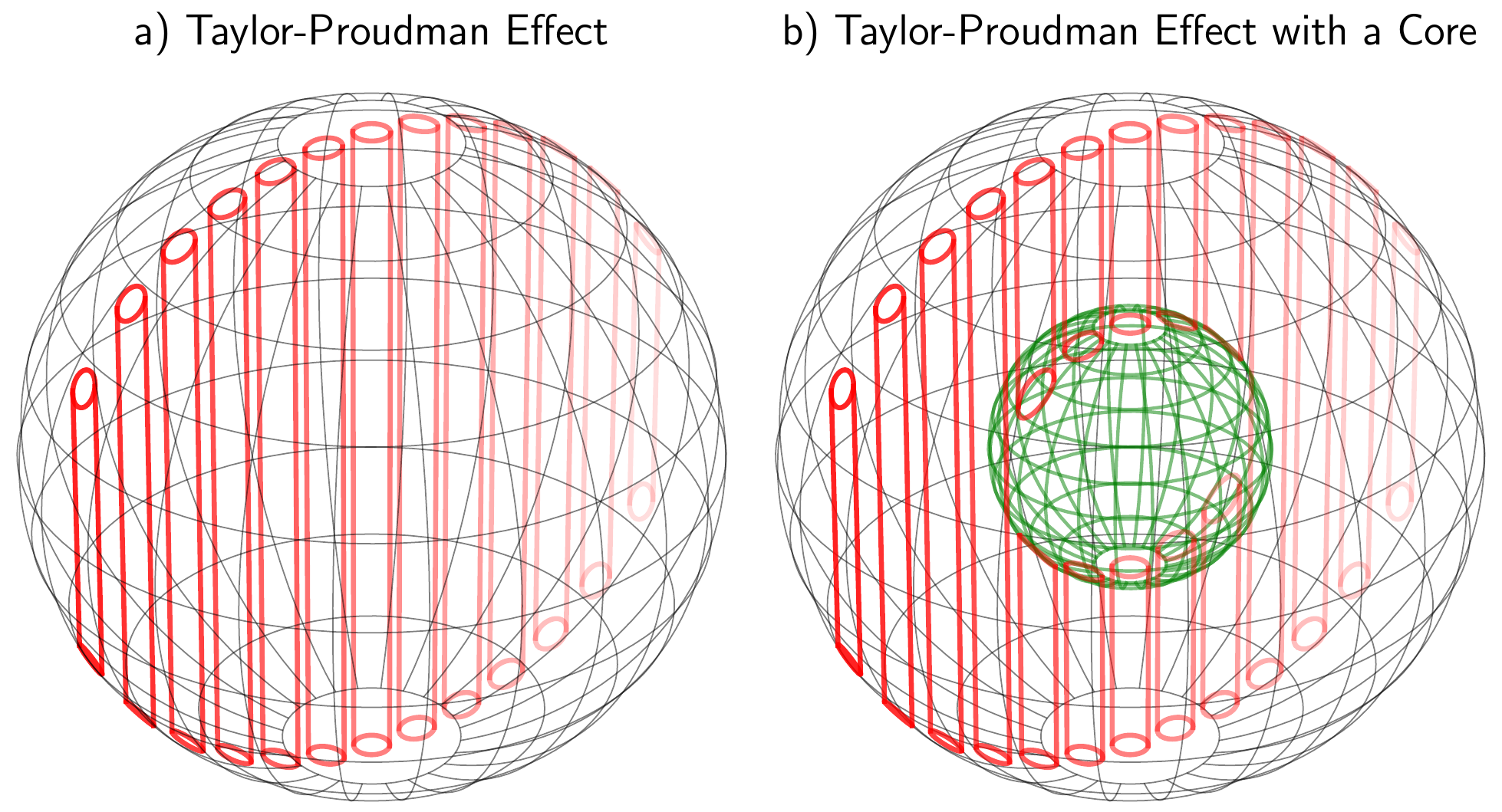}
\caption{An illustration of the Taylor--Proudman effect of a rotating liquid
sphere (black), with and without a core (green). The liquid tends to organize into
rotating cylinders (red), with each cylinder rotating in an opposite sense
compared to the one next to it. While cylinders occupy the entire volume of the
sphere, only a cross-section is shown here for clarity. The presence of a core
will disconnect the cylinders that emanate at high latitudes. This is known as the
\textit{deep model} of atmospheric dynamics.}
\label{fig:taylorproudman}
\end{figure}

\subsection{Circulation Cells}
\label{sec:circulation}

The banded structure of Jupiter is nontrivial to explain, and it is helpful to
first consider Earth. On Earth, the \textit{Hadley cell} features upwelling from
the equator and downwelling at midlatitudes ($\sim$30\degr\ latitude), forming a
closed loop driven by solar heating, known as \textit{thermally direct
circulation}. At around 60\degr\ latitude, warm air rises and is transported
toward the pole, where it cools and sinks, completing the loop. Between the Hadley
cell and the \textit{polar cell} sits the \textit{Ferrel cell}, whereby part of
the air driven upward at 60\degr\ moves equatorward until it encounters the
downwelling of the equatorial Hadley cell. While a similar stack of circulation
cells could be generating the banded structure of Jupiter, the scenario is made
more complicated by the fact that the planet lacks a solid surface and spins much
more rapidly.

There are two basic models that describe how the observed zonal winds develop
with depth. The first model describes the weather observed at the top of the
atmosphere as \textit{shallow}, which is closely analogous to that on Earth, where
the winds are naturally bound by a lower boundary (i.e., the surface). At the
giant planets, such a boundary would move at an angular velocity equal to the
interior rotation period, and there could exist strong vertical velocity
gradients within the shallow weather layer. While this model works well for the
terrestrial planets, it has several shortcomings at the giant planets: (a) In
order to maintain a banded structure that remains largely invariant over long
timescales, energy is required to be pumped into the system. (b) These models
predict that the midlatitude circulation cells are the same size as the equatorial
cell, which is not the case. (c) They predict strong westward equatorial jets at
the gas giants (Jupiter and Saturn), whereas the jets are observed to be eastward.

The second model is the \textit{deep model}, where the observed wind at the top of
the troposphere extends deep into the interior. According to the Taylor--Proudman
effect \citep{1923RSPSA.104..213T}, a rapidly rotating fluid sphere will organize
into many rotating concentric cylinders that are aligned along the rotational axis
\citep{1982Icar...52...62I}. This is illustrated in
Figure~\ref{fig:taylorproudman}. In this scenario, the belts and zones at low to
midlatitudes are physically connected to their counterparts in the opposite
hemisphere via these cylinders, which would explain the symmetric nature of
Jupiter's banded structure. At higher latitudes, however, the presence of a
liquid helium core intersects the cylindrical flow, and so the symmetric banded
structure would break down in these regions and create a more stochastic
atmosphere. The boundary between the two regimes of connected and disconnected
cylinders is known as the \textit{critical latitude}. This simple idea largely
agrees with what is observed, particularly on Jupiter. The size of these
cylinders, and therefore the resultant banded structures of a planet, are
governed by both the size and rotation rate, and smaller planets with a longer
rotation rate generate larger cylinders and therefore fewer belts and zones. The
deep model predicts that the zonal winds observed at the top of the cloud deck
would continue very deep into the atmosphere, which was indeed observed by the
Galileo probe down to the 240-atmosphere pressure level \citep{2003NewAR..47....1Y}.

\subsection{Turbulence and Instabilities}
\label{sec:turbulence}

The basic circulation pattern outlined in Section~\ref{sec:circulation} provides
an understanding of the broadscale movement of the atmosphere. However, superposed
on this are smaller-scale movements that depart from the mean flow, known as
\textit{eddy motions}. These take the form of \textit{turbulence, vortices}, and
\textit{waves}, which arise from instabilities produced by gradients in
temperature, motion, or pressure. Turbulence is ultimately a consequence of
instabilities in the atmosphere, and the \textit{Richardson number} provides a
measure of the stability of turbulence; if the number is less than one, the
turbulent behavior will persist. There are several instabilities relevant to the
atmospheres of the giant planets:

\begin{itemize}
\item \textit{Static instability} occurs when a rising parcel cools more quickly
than the surrounding atmosphere (i.e., the adiabatic lapse rate), which in turn
makes it less buoyant, disrupting the convective flow.

\item \textit{Inertial (or dynamic) instability} occurs when the pressure gradient
force and the centrifugal forces are not evenly balanced inside a vortex, and the
frictional forces can generate heat that drives instabilities.

\item \textit{Barotropic instability} happens in a regime where the vertical
temperature across a region remains constant along lines of constant pressure
(which is the barotropic condition). Here, instabilities can arise from a shear in
the horizontal velocity field, which in turn can generate vortices.

\item \textit{Baroclinic instability} occurs in conditions where the barotropic
condition does not hold (i.e., temperatures do not remain constant along lines of
constant pressure). Consequently, instabilities can be generated by changing
vertical temperature gradients.

\item \textit{Kelvin--Helmholtz instability} is produced when two adjacent regions
have different wind speeds, creating a shear between the two.
\end{itemize}

\subsection{Waves}

Different types of waves play important roles in the atmospheres of the giant
planets. They can alter the temperatures and densities along their path, and they
can transfer energy and momentum from one atmospheric layer to another or within
an atmospheric layer. Waves are simply characterized by their amplitude,
wavelength, and phase, which can evolve over time. At the giant planets, there are
three types of waves that are particularly important:

\textit{Gravity waves} (not to be confused with the astrophysical ``gravitational
waves'') are generated at the boundaries of atmospheric parcels with different
motion (e.g., in turbulent regions). Gravity waves, generated in the troposphere,
are divided into two categories: high-frequency \textit{internal} gravity waves
and low-frequency \textit{inertia} gravity waves. These can propagate both
vertically and horizontally. When they travel upward in altitude, they can break
in the stratosphere and upper atmosphere, coupling these different regions. In the
stratosphere, these waves provide the primary energy source for eddy mixing
\citep{1985JGR....9013067S}. These waves can also reach the upper atmosphere, and
fine-scale ionospheric disturbances observed above the GRS on Jupiter were linked
to the deposition of multiple gravity waves \citep{2024NatAs.tmp..120M}.

\textit{Rossby waves} have long wavelengths and large amplitudes, undulating
horizontally along lines of constant longitude in rotating systems. They are a
type of inertial wave and arise from the conservation of potential vorticity. On
Earth, Rossby waves are believed to be responsible for the quasibiennial
oscillation, and on Jupiter, a similar phenomenon is observed---the
quasiquadrennial oscillation, which manifests as a temperature perturbation in the
stratosphere that slowly moves down in altitude with a period of about 4 years
\citep{1991Natur.354..380L}.

The Kelvin--Helmholtz instability outlined in Section~\ref{sec:turbulence} is
generated at the interface between two different flow regimes. If the shear flow is
large enough, \textit{Kelvin--Helmholtz waves} form along the boundary, and they
provide a mechanism for transferring energy and momentum between the two regions.

\subsection{Lightning}

Electrification is the build-up of charge within a parcel of the atmosphere and is
a common process. On Earth, clouds contain water in different phases (i.e.,
liquid, solid, and ice). Light ice crystals are transported upward in altitude in
the convective core of a cloud, while larger and heavier soft hail, known as
\textit{graupel}, moves downward. Collisions between them transfer electrons to
the heavier particles, rendering the top of the cloud positively charged and the
bottom negatively charged.

Lightning is the electrostatic discharge between two regions with different
electric charges that have undergone electrification. Its physical manifestation
is a bright flash caused by electrons tearing through the atmosphere, followed by
thunder associated with the shock wave produced by the movement of the electrons,
producing compressional waves in the atmosphere. Lightning has been directly
observed on Jupiter \citep{1979Natur.280..794C} and Saturn
\citep{1981Sci...212..239W} and has been indirectly detected on Uranus
\citep{1986Natur.323..605Z} and Neptune \citep{1990JGR....9520967G}. This process
is believed to be similar to what is seen on Earth, generated in the low-altitude
water clouds. Instead of the terrestrial graupel, the term \textit{mushball} has
been used, where soft hail is enriched with ammonia
\citep{guillot2020a}.

\subsection{Disequilibrium Species}

For a particular molecular species, there is a temperature and pressure at which
it can exist in chemical equilibrium, where the molecular abundance remains
constant and the production and destruction chemical reactions are equally
balanced. Molecules can also exist in a disequilibrium state, whereby the
destruction or production processes will dominate over the other, leading to
reduced or increased abundances relative to their equilibrium state. Molecules are
transported vertically from altitudes at which they are in equilibrium to a state
of disequilibrium by the strong overturning of the dynamic atmosphere. Therefore,
observing species in a state of disequilibrium will provide essential information
about the dynamics and circulation of the atmosphere. At the giant planets,
important disequilibrium species are nitrogen, carbon, germane, arsine, ammonia,
and phosphine, as well as different spin states of atoms in the hydrogen molecule.

\begin{figure}[ht]
\centering
\includegraphics[width=0.8\linewidth]{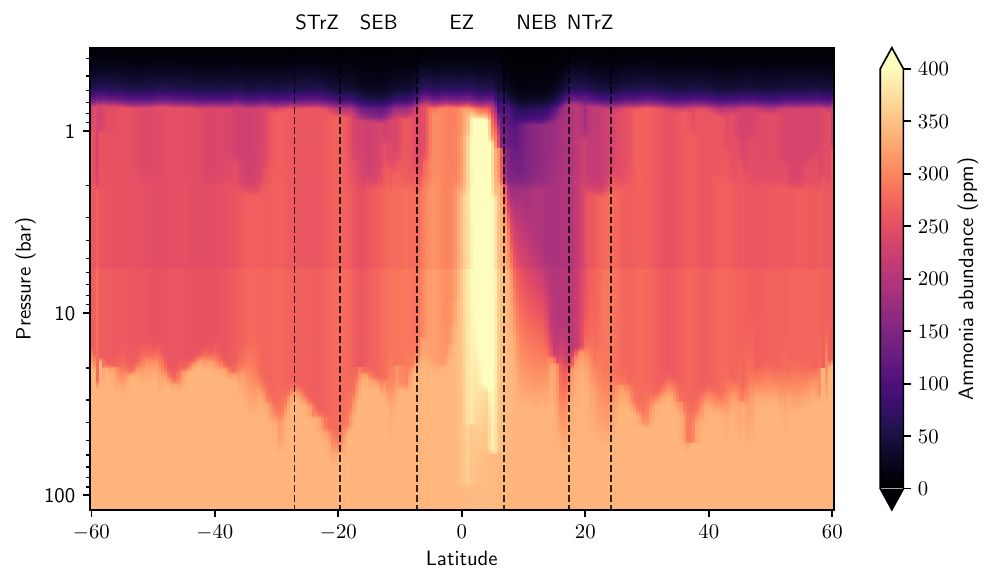}
\caption{The deep vertical ammonia abundance as measured by the Juno Microwave
Radiometer (MWR). \textit{Source}: Adapted from \citet{2023PSJ.....4...25M}.}
\label{fig:ammonia}
\end{figure}

The first observation of how a disequilibrium species is distributed at great
depth on Jupiter came from the Juno MWR, which observed the deep abundance of
ammonia \citep{2017GeoRL..44.5317L,2023PSJ.....4...25M}, as shown in
Figure~\ref{fig:ammonia}. It shows significant variations in the abundance at high
pressures ($>$10~bar) and a significant enhancement throughout the entire
atmospheric column at the equator, with a depletion just north of this
enhancement. This curious distribution may be related to the presence of
thunderstorms, where the formation of mushballs efficiently removes ammonia from
that column of atmosphere. The absence of thunderstorms at the equator then allows
an enrichment in the Equatorial Zone \citep{2020JGRE..12506404G}.

\section{Gas Giants}
\label{sec:gasgiants}

Jupiter and Saturn are classed as gas giants because their bulk composition
consists mainly of hydrogen and helium, and this composition makes them different
from the ice giants, Uranus and Neptune.

\subsection{Jupiter}

The visual appearance of Jupiter is dominated by the bright \textit{zones} and
dark \textit{belts}, which appear in pairs across the planetary disk. The zones
are regions in which convective upwelling occurs, bringing material from the deep
atmosphere, such as condensable and disequilibrium species, and the subsequent
downwelling occurs in the belts. As a result of this convection, the zones are
hotter than the belts, since deeper tropospheric altitudes experience lower
temperatures. There is a strong east--west (i.e., zonal) wind shear between the
belts and the zones, creating instabilities (see Section~\ref{sec:turbulence})
that can produce vortices or cyclonic storms. The zones closest to the Equatorial
Zone (EZ) are called the North and South Equatorial Belts (NEB and SEB), followed
by the North and South Tropical Zones (NTrZ and STrZ). Apart from the turbulent
behavior at the boundaries of these regions, they appear stable on timescales of
decades. However, they routinely undergo dramatic changes, some of which are
cyclic \citep[e.g.,][]{2018GeoRL..4510987A}.

The belt-zone structure is broadly compatible with the circulation patterns
outlined in Section~\ref{sec:circulation}, with upwelling at the EZ and
downwelling in the Equatorial Belts, forming a Hadley cell, followed by belts and
zones produced by the deep Taylor--Proudman circulation. However, observations by
the Juno microwave radiometer, which penetrates deep into Jupiter's troposphere,
have hinted that this simple single-cell behavior may be more complex. Instead of
one cell, Jupiter may have stacked circulation cells, whereby at depth, the EZ
experiences downwelling \citep{2021JGRE..12606858F,2000Natur.403..630I}. The
boundary at which the cylinders switch direction has been termed the
\textit{jovicline}. Toward the poles, above latitudes of $\sim$70\degr, this
patterned behavior is replaced with a more chaotic, asymmetric atmosphere, and
stable circumpolar cyclones have been detected by the Juno spacecraft in both the
north and the south \citep{adriani2018}.

\begin{figure}[ht]
\centering
\includegraphics[width=0.8\linewidth]{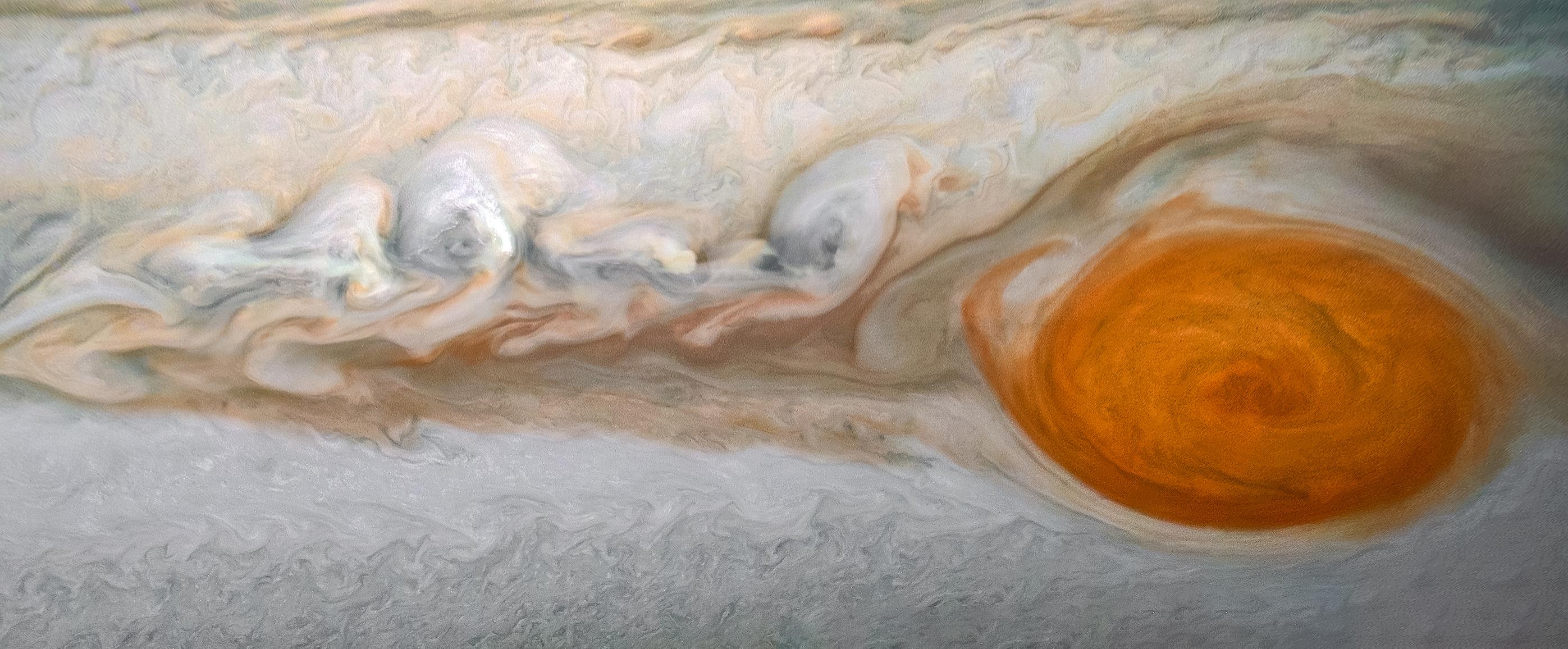}
\caption{A JunoCam observation of Jupiter's Great Red Spot (GRS) from September 7,
2023. \textit{Source}: NASA/JPL-Caltech/SwRI/MSSS, image processing by Kevin
M.~Gill, \textcopyright\ CC BY.}
\label{fig:grs}
\end{figure}

The most striking feature in the atmosphere of Jupiter is the GRS, as shown in
Figure~\ref{fig:grs}. It is the largest and longest-lived vortex in the Solar
System. The center of the storm shows large amounts of aerosol reflectance and
relatively modest wind speeds. The bulk of the motion is confined to a ring
surrounding the storm that prevents any mixing with the surrounding regions
\citep{2010Icar..208..306F}. The storm is relatively shallow, extending only about
500~km in altitude \citep{2021Sci...374..964P}, which means that its vertical
extent is only 0.5\% of its horizontal extent and that it is disconnected from the
deep zonal flows that surround it. The distinct reddish color of the storm is
thought to be due to some hitherto unknown chromophore, likely a photochemical
product of ammonia and acetylene \citep{2019Icar..330..217B}.

\subsection{Saturn}

The international Cassini mission to Saturn spent 13 years orbiting the planet,
completing 294 orbits. This hugely successful mission made many significant
discoveries, including cryovolcanic geysers on the moon Enceladus
\citep{2006Sci...311.1406D}, methane lakes on Titan \citep{2007Natur.445...61S},
and the fact that measurements of the rotation rate appear to yield different
results in the northern and southern hemispheres \citep{2009GeoRL..3616102G}.
Additionally, Cassini carried the European Huygens probe, which became the first
probe to land on a moon in the outer Solar System \citep{2009AARv..17..149L}.

Saturn has an axial tilt of $\sim$27\degr, which produces strong seasonal
behavior. The changing solar illumination alters the stratospheric photochemistry,
and at the northern solstice, increased absorption of sunlight by methane drives
increased temperatures in the upper troposphere and lower stratosphere
\citep{2018NatCo...9.3564F}.

\begin{figure}[ht]
\centering
\includegraphics[width=0.6\linewidth]{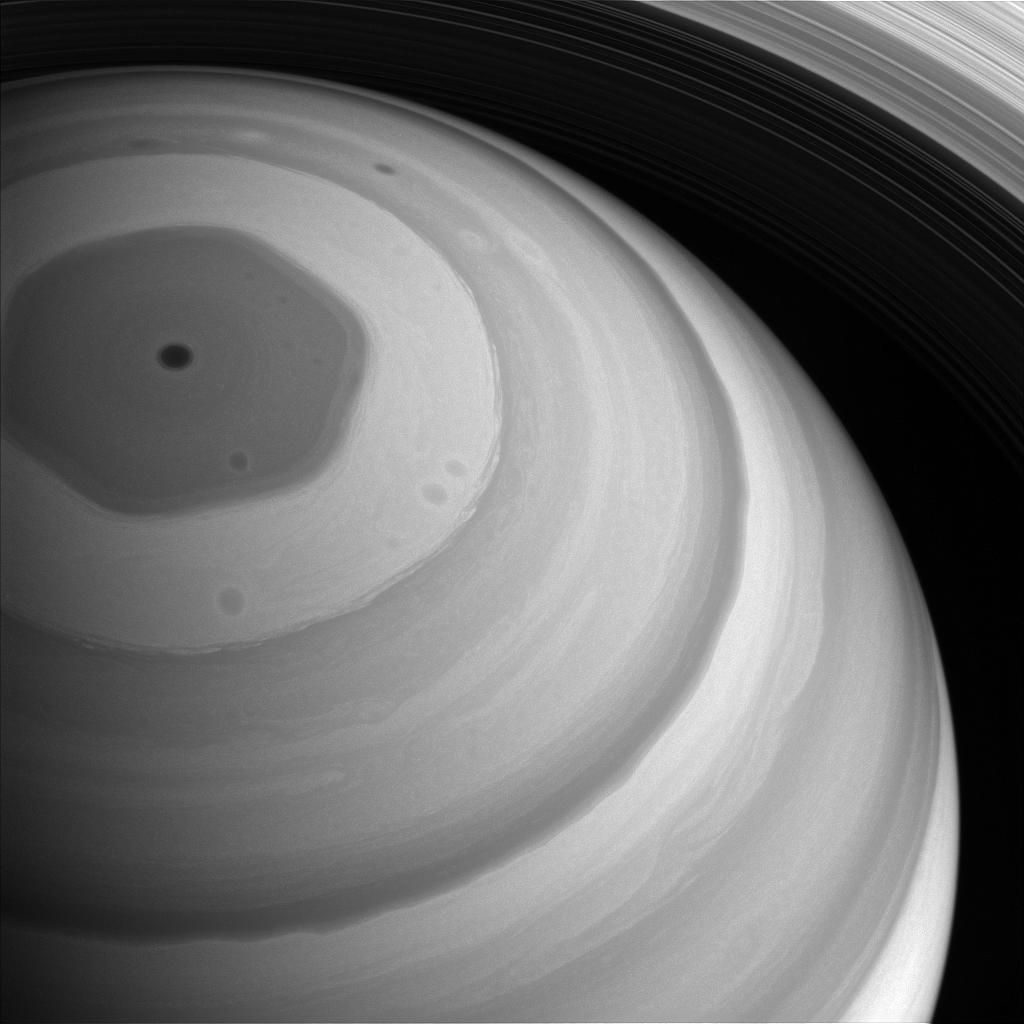}
\caption{The hexagon in Saturn's northern hemisphere obtained by Cassini at
728~nm. \textit{Source}: NASA/JPL-Caltech/Space Science Institute.}
\label{fig:hexagon}
\end{figure}

The tropospheric zonal wind speeds at Saturn show a similar intricate structure to
that of Jupiter (Figure~\ref{fig:zonalwinds}), also associated with broadscale
belt and zone structures (Figure~\ref{fig:hexagon}). Saturn exhibits much larger
velocities than Jupiter, especially at the equator, where winds can reach up to
400\msinv, and these are likely linked to the deep motion of the atmosphere in the
form of Taylor--Proudman-type cylinders. Using radio observations from the ground,
the stratospheric wind speeds can also be measured, showing that the equatorial
jet extends up to the 10~$\mu$bar level \citep{2022AA...666A.117B}.

One of the more remarkable features of Saturn's troposphere is the presence of a
persistent polar hexagon in the northern hemisphere \citep{1988Icar...76..335G},
as shown in Figure~\ref{fig:hexagon}. This six-sided feature is very stable and
remains fixed in longitude \citep{1990Sci...247.1206G} and is believed to be
formed by a Rossby wave generated by instabilities in the nearby zonal jet at
78\degr N \citep{1990Sci...247.1061A}. The Cassini spacecraft acquired remarkable
views of this feature, and midinfrared observations show that it extends well up
into the stratosphere, altering the temperature structure
\citep{2018NatCo...9.3564F}.

A semiregular occurrence at Saturn is the emergence of an intense convective
bright storm in the northern spring \citep{2011JBAA..121..270M} that eventually
extends and envelops the northern hemisphere. The instability that generates the
intense moist convection of these events is likely linked to the fact that the
stratospheric temperature is strongly driven by the seasonal illumination, but the
preference for the northern hemisphere remains unknown
\citep{2018stfc.book..377S}.

\section{Ice Giants}
\label{sec:icegiants}

The ice giants, Uranus and Neptune, are smaller and denser than the gas giants
(see Table~\ref{tab:planets}). While the outer envelopes of these planets are
dominated by hydrogen and helium, they have a higher proportion of heavier
elements at depth. The term \textit{ice giants} derives from the fact that, at
depth, they may well be dominated by ``icy'' materials, such as water and methane,
though in a hot and fluid state \citep{1995netr.conf..109H,1991uran.book...29P}.
However, considerations of the distribution of carbon monoxide and its sources
(i.e., external or internal) have led to the suggestion that the interiors may be
more dominated by rock than by ices, and perhaps the term \textit{rock giant}
would be more appropriate \citep{2020RSPTA.37890489T}.

The atmospheres of these planets are very cold (see Figure~\ref{fig:tprofiles}),
which means that the thermal emission features from their atmospheric constituents
emit weakly, and their distance from the Sun also renders reflected light very
weak, with both effects making them challenging to observe. Uranus and Neptune
have only been visited by a single spacecraft---both were visited by the Voyager 2
spacecraft in 1986 and 1989, respectively. Hence, scientific understanding of
these two planets is much more limited than that of Jupiter and Saturn. Thus, it
is certain that a new interplanetary spacecraft mission to either of these planets
would provide invaluable information about these worlds.

\begin{figure}[ht]
\centering
\includegraphics[width=0.6\linewidth]{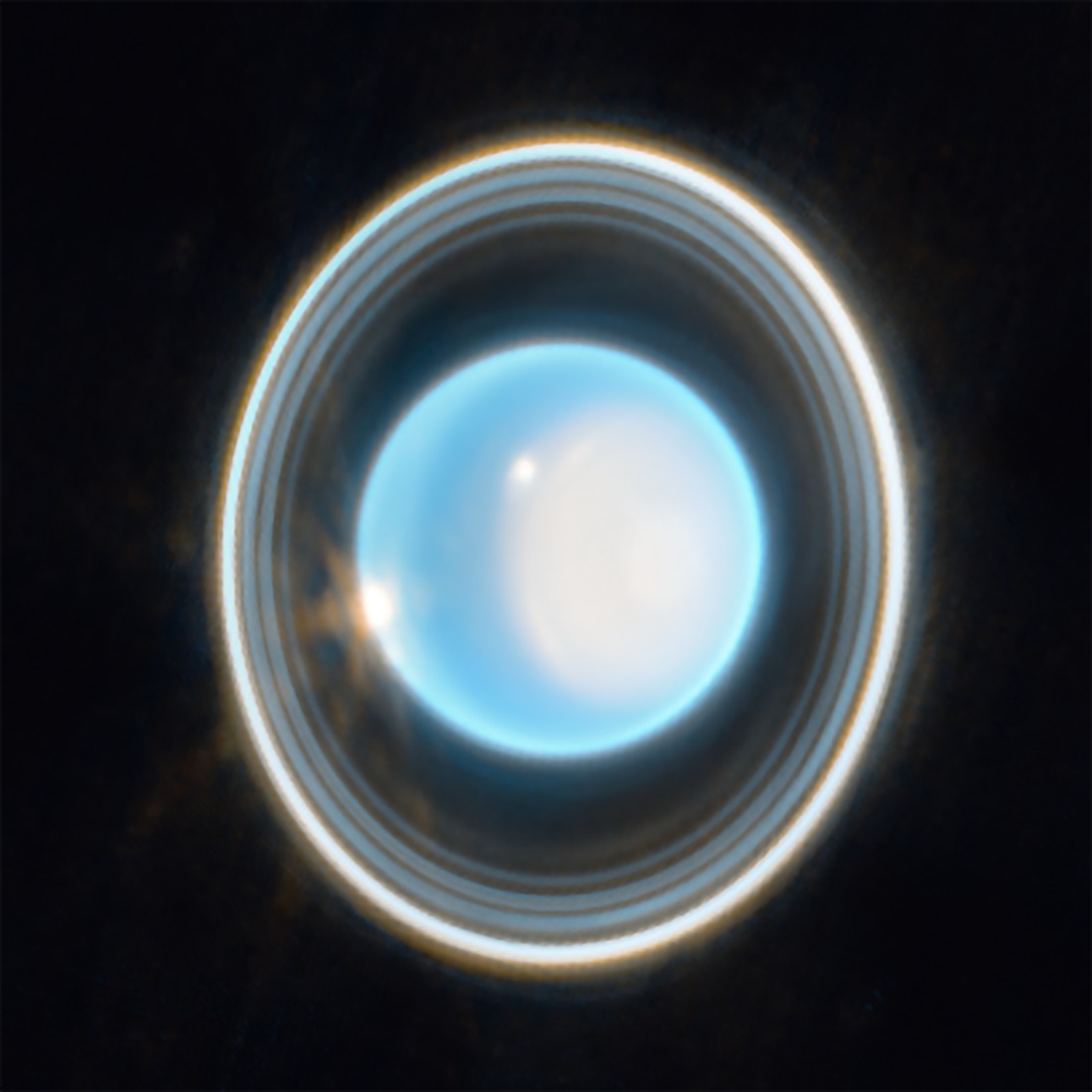}
\caption{Uranus imaged by NIRCam on the James Webb Space Telescope (JWST) in the
infrared. \textit{Source}: NASA, ESA, CSA, STScI.}
\label{fig:uranus}
\end{figure}

\subsection{Uranus}

Uranus is unique in many ways. It effectively rotates ``on its side,'' generating
extreme seasons, during which the northern pole is sunlit all the time at northern
solstice (Figure~\ref{fig:uranus}). The planet also has very little internal heat
\citep{1990Icar...84...12P}. Models have suggested that the rotational geometry and
lack of internal heat could be linked: a large collision in the early Solar System
could have knocked the planet on its side and at the same time released most of its
stored internal heat \citep{2020NatAs...4..880I}. The strange nature of Uranus's
orbital geometry makes it an interesting laboratory for understanding planetary
atmospheres.

The lack of internal heat should, in principle, produce a very sluggish
atmosphere, with the lack of energy available at the bottom of the troposphere
severely limiting convection. However, there are several indications that despite
the lack of an internal heat source, significant dynamics are present on the
planet. First, with limited internal heat, solar forcing may be expected to
dominate as a heat source, which would produce a warmer atmosphere at the subsolar
point and a cooler atmosphere toward the limb. However, Voyager 2 measured very
little variability in the temperature at the top of the troposphere
\citep{1987JGR....9215011F,1986Sci...233...70H}. Second, just like on Jupiter,
condensable species on Uranus, such as methane and hydrogen sulfide, are important
tracers of tropospheric circulation. These are observed to have strong gradients
indicative of equator-to-pole circulation that appears independent of solar
irradiation \citep{2020RSPTA.37890477M}. Third, the extreme and long seasons on
Uranus should generate strong seasonal behavior if sunlight dominates as an energy
source. This is not observed, however, and the structure of the temperatures in
the upper troposphere and lower stratosphere has remained largely constant over
the 40 years since the Voyager 2 flyby
\citep{2015Icar..260...94O,2020AJ....159...45R}.

The zonal wind speeds of Uranus in Figure~\ref{fig:zonalwinds} show retrograde
winds at the equator and prograde winds at midlatitudes. Analysis of the
higher-order moments of the gravity field suggests that these winds are relatively
shallow, extending 1,000~km in altitude \citep{2013Natur.497..344K}. In the
troposphere, the broadscale circulation is believed to be upwelling at the equator
and downwelling at the pole \citep{1998Icar..135..501C}, painting a much simpler
picture than what is seen on Jupiter and Saturn. However, Uranus does not lack
complexity. For example, Uranus is likely to have intricate cloud structures where
cold temperatures allow methane to condense to form clouds in the troposphere
\citep{1987JGR....9214987L}, along with clouds of water, hydrogen sulfide, and
ammonia hydrosulfide \citep{2020RSPTA.37890476H}.

In the stratosphere of Uranus, methane photochemistry drives heating, just like on
Jupiter. However, the limited internal heat limits the vertical mixing by eddy
diffusion, which produces a relatively deep methane homopause
\citep{1987JGR....9215093H,2018Icar..307..124M}, which in turn produces an extended
upper atmosphere \citep{2020RSPTA.37890478M}.

\subsection{Neptune}

\begin{figure}[ht]
\centering
\includegraphics[width=0.6\linewidth]{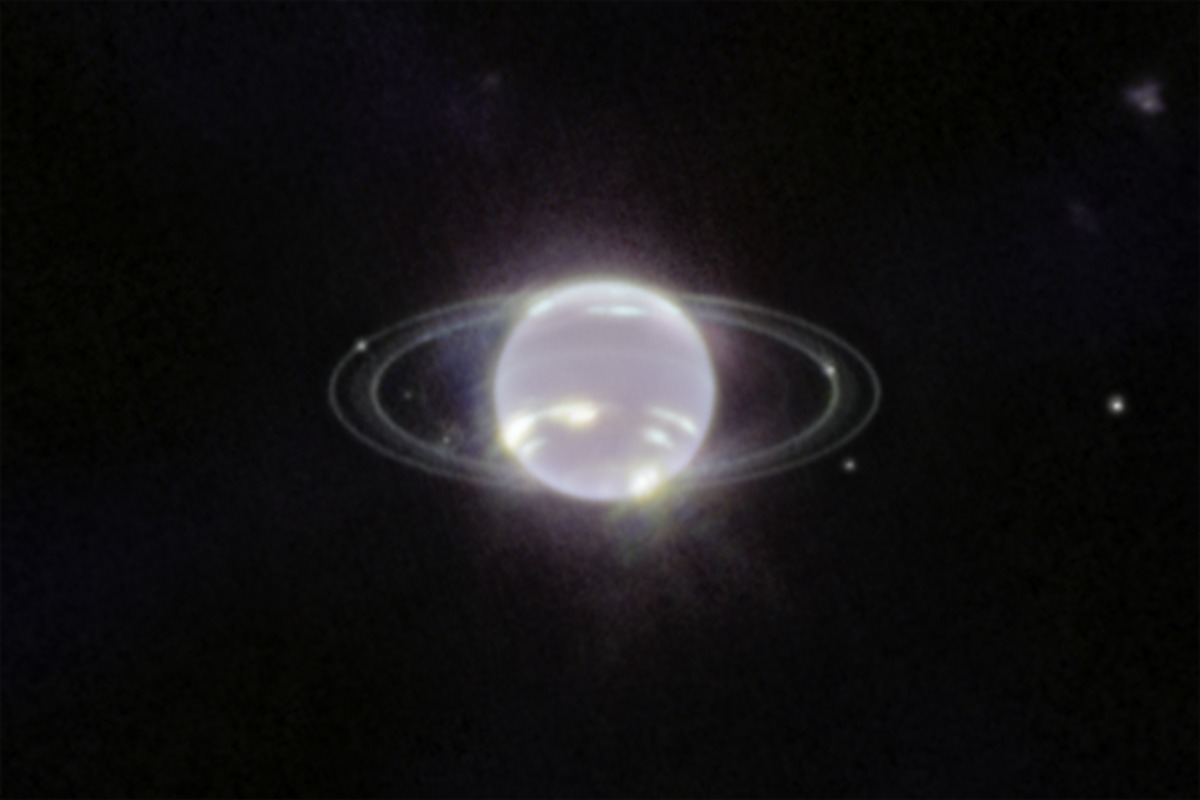}
\caption{Neptune imaged by NIRCam on the JWST. \textit{Source}: NASA, ESA, CSA,
STScI, image processing by Joseph DePasquale (STScI) and Naomi Rowe-Gurney
(NASA-GSFC).}
\label{fig:neptune}
\end{figure}

The visual appearance of Neptune is dominated by bright clouds that disappear
within days of appearing (Figure~\ref{fig:neptune}). The rapidly changing nature
of these outbreaks is likely primarily driven by the large internal heat flux of
the planet, which drives strong moist convection in the lower atmosphere. A similar
complex set of cloud decks as is expected on Uranus is likely present on Neptune.

The tropospheric zonal winds of Neptune are the fastest observed among the giant
planets (Figure~\ref{fig:zonalwinds}). The equator is strongly retrograde with wind
speeds of up to 400\msinv, with weaker prograde winds of $\sim$250\msinv\ at higher
latitudes. The planet is thought to have a similar tropospheric convection pattern
to that of Uranus, with upwelling at the equator and downwelling at the poles.
Using radio observations, the stratospheric winds can be directly measured
\citep{2023AA...674L...3C} and decay significantly with altitude, in line with what
is derived by applying the thermal wind equation to the tropospheric winds
\citep{2014Icar..231..146F}. As the winds decay with altitude, a drag force is
inferred, which may be in the form of breaking gravity waves. Ultimately, the
mechanisms that drive these extreme and largely unchanging winds remain unknown.

While the lower stratospheric temperature of Uranus seems to change very little
with time, the stratosphere of Neptune can change on timescales much shorter than
the long seasons \citep{2022PSJ.....3...78R}. These changes may in part be related
to the 11-year solar cycle, which modulates the solar ultraviolet emissions, which
in turn regulate stratospheric photochemistry. There is also a solar cycle
dependence on the emergence of bright clouds \citep{2023Icar..40415667C}, which may
indicate that even at a great distance from the Sun, it still governs fundamental
aspects of Neptune's atmosphere.

The large internal heat flux of Neptune drives vigorous eddy diffusion in the
stratosphere, which, in turn, has the effect of producing a very high methane
homopause \citep{2018Icar..307..124M}, which in turn produces a very compact
thermosphere and ionosphere, in stark contrast to Uranus, whose methane homopause
and distribution of hydrocarbon daughter products are very shallow
\citep{2014Icar..243..471O}.

\section{Tools of the Trade}
\label{sec:tools}

Most understanding of the atmospheres of the giant planets is derived from remote
sensing observations, where the planets are observed at great distances. The one
exception to this is the Galileo Probe, which provided in situ measurements and is
discussed later in this section.

The atmospheres of these planets can be explored both with imaging and
spectroscopy targeting different wavelength regions. Since the observed radiances
from molecular bands are dependent on temperature, separate observations are often
obtained at wavelength regions from which the temperature can be determined. For
example, to investigate the abundances of the methane photochemical products
acetylene and ethane in the stratosphere of Jupiter, which can be observed at
13.4\um\ and 12.2\um, respectively, the temperature at these altitudes can be
constrained by observing the methane emission at 8.0\um\
\citep[e.g.,][]{2018Icar..305..301M}. The midinfrared contains a plethora of
emission and absorption features from giant planet atmospheres and is therefore
commonly used, but other wavelength bands also contain important information, from
the ultraviolet to the microwave.

\subsection{Ground-Based Telescopes}

Telescopes on the ground provide important constraints on the atmospheres of the
giant planets. However, they are limited by Earth's atmosphere, which effectively
sits between the telescope and the planet in question, in two basic ways. First,
the turbulent motion of the telluric atmosphere introduces a blurring of the
observations, known as \textit{seeing}. This can be corrected by using
\textit{adaptive optics}, available at some telescopes, which correct for this by
deforming the primary or secondary mirror at a very high rate. Second, the
molecules in Earth's atmosphere absorb light at certain wavelengths, making certain
parts of the electromagnetic spectrum inaccessible from the ground. For example,
water, carbon monoxide, and methane absorb across broad bands in the near and
midinfrared. Further complexity is added if one studies these species at the giant
planets: the planets must be observed at an appropriate time when the relative
motion between the Earth and the planet effectively Doppler shifts these lines away
from the molecular features in Earth's atmosphere.

Most of the world's premier observatories are found at high altitude, above Earth's
tropospheric inversion layer. For example, Mauna Kea in Hawaii is an ideal site for
infrared astronomy because of its dry and stable atmosphere, sitting 4,100~m above
sea level.

\subsection{Space Telescopes}

By placing telescopes in space, the limitations introduced by observing through
Earth's atmosphere can be eliminated. The first operational space telescope was the
Orbiting Astronomical Observatory-2, launched in 1968, which operated until 1973.
Since then, there have been a number of space telescopes, but arguably the most
important is the Hubble Space Telescope, launched in 1990. Long-term programs like
the Outer Planet Atmospheres Legacy program have provided invaluable monitoring of
the four giant planets since 2015.

The James Webb Space Telescope (JWST) was launched on Christmas Day 2021 and is the
largest and most complex infrared observatory ever constructed. It orbits far away
from Earth (unlike the Hubble Space Telescope) and was designed to sense the first
stars and galaxies formed in the very early universe. Therefore, it has four
extremely sensitive instruments, covering a wavelength range between 0.6 and
28.5\um\ for both imaging and spectroscopy. All four of the giant planets have been
observed, and the early results have revealed Jupiter's Great Red Spot in great
detail \citep{2024JGRE..12908415H}, narrow jets in Jupiter's equatorial
stratosphere \citep{2023NatAs...7.1454H}, complex patterns in Jupiter's ionosphere
\citep{2024NatAs.tmp..120M}, and seasonal changes in Saturn's polar atmosphere
\citep{2023JGRE..12807924F}. This is only the advent of exploring the giant planets
with JWST, and as more data are gathered, especially at the ice giants where the
molecular features are very weak, scientists are poised to gain a much deeper
understanding of these systems.

\subsection{Spacecraft Missions}

Even with the best telescopes in the world, the finer details of atmospheric
structure and dynamics can only be resolved by spacecraft flying close to the
planet. The first close-up views of Jupiter in the visible and infrared were
provided by Pioneer 10 in 1973, followed by Pioneer 11 in 1974, with the latter
reaching Saturn in 1979. These missions were the preamble to the Voyager 1 and 2
Grand Tour, with the Voyager 2 spacecraft going on to visit all four giant planets,
thanks to a fortuitous planetary alignment that only occurs once every 175 years.
Voyager 1 flew by Jupiter in 1979 and Saturn in 1980 and included a flyby of Titan,
Saturn's largest moon. The orbital constraints required for the Titan flyby took the
probe out of the orbital plane. Voyager 2 went on to visit both Uranus in 1986 and
Neptune in 1989. These bold missions laid the foundation for the modern-day
exploration of the giant planets, and their importance cannot be overstated.
Amazingly, both Voyager 1 and 2 are still operational, exploring the heliospheric
boundary.

The Galileo mission to Jupiter launched aboard Space Shuttle Atlantis in 1989,
following a five-year delay caused by the Challenger disaster. The mission was
severely hampered by the fact that the high-gain antenna did not deploy properly,
probably related to the degradation of lubricants during the extended period that
the spacecraft sat in storage. Nonetheless, the spacecraft delivered the Galileo
Entry Probe upon arrival in 1995 and deorbited in 2003. Despite the limited data
rates that Galileo could transmit back to Earth, the mission achieved most of its
science objectives: besides being the first to deliver a direct probe of an
outer-planet atmosphere, it was the first to make spectral images of Jupiter in the
near-infrared.

NASA's Juno mission, launched in 2011, entered orbit around Jupiter in 2016. The
highly inclined orbits pass above the poles and skim the top of the equatorial
atmosphere during each pass. The prime mission was highly targeted to study the
auroral process and the magnetic and gravity fields of Jupiter. The mission was, as
of 2025, in its extended phase, which continues its close flybys of Jupiter, as
well as flybys of Jupiter's moons.

The very successful Cassini mission to Saturn, which also flew by Jupiter in late
2000, set a gold standard for international collaboration. These large flagship
missions are very expensive and complex, and by involving expertise and funding
from multiple space agencies, the considerable challenges of these projects can be
solved. This is clearly good for the scientific outcome of these missions, but it
also brings international communities together.

\begin{figure}[ht]
\centering
\includegraphics[width=0.8\linewidth]{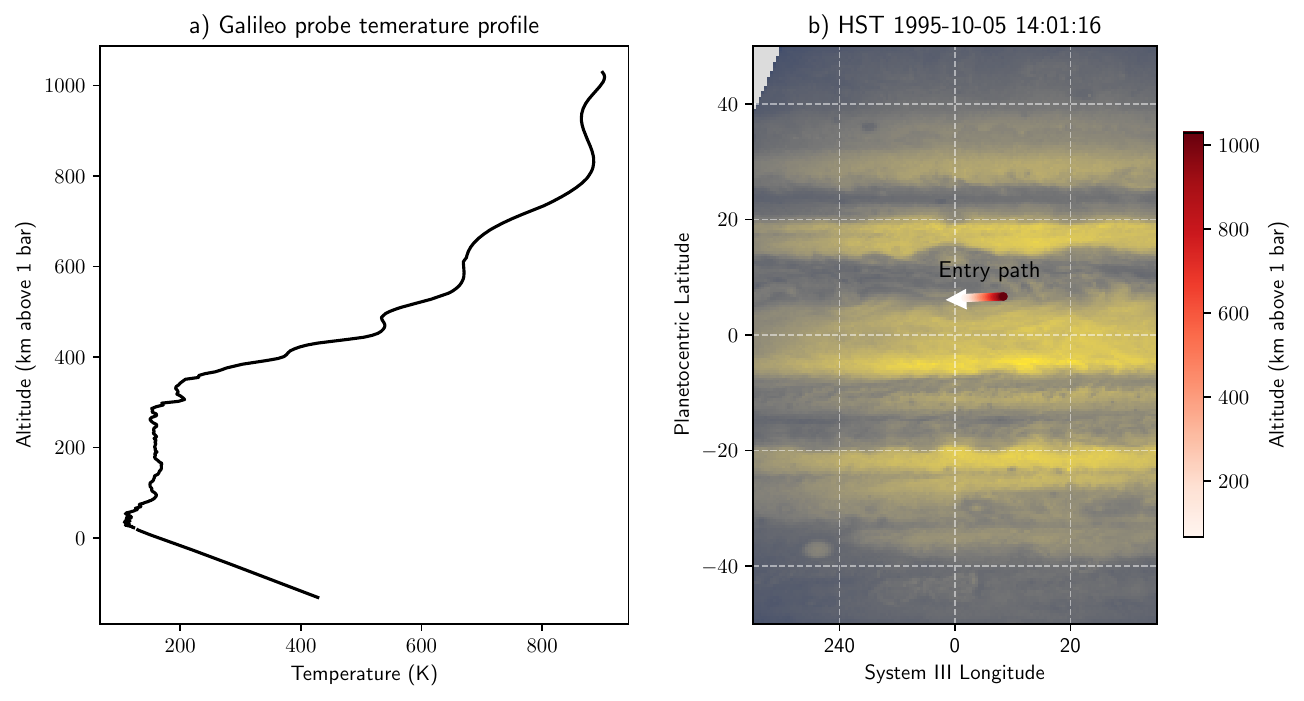}
\caption{(a) The vertical temperature profile as measured by the Galileo probe
\citep{1996Sci...272..844S} on December 7, 1995. (b) A Hubble Space Telescope Wide
Field Camera 3 image taken on October 5, 1995, using the F410M filter. The entry
path of the probe while traversing the upper atmosphere is shown, where the color
indicates the altitude of the spacecraft measurement.}
\label{fig:galileo}
\end{figure}

\subsection{Entry Probes}

There are distinct advantages to sending a probe into an atmosphere to measure the
composition, temperature, and pressure as it descends. First, it provides a
``ground truth'' for remote sensing observations. Second, there are certain
important species that cannot easily be measured via remote sensing at high
atmospheric pressures. In particular, the noble gases (helium, neon, argon,
krypton, and xenon) provide critical constraints on the formation and evolution of
these planets.

The only in situ observations of a giant planet atmosphere came from the Galileo
probe, which entered Jupiter's atmosphere on December 7, 1995, and descended to
pressures of 24~bar before the signal was lost. It carried a suite of instruments
that measured the vertical composition and temperature structures.
Figure~\ref{fig:galileo} shows the measured vertical temperature profile along with
a reference HST WFC3 image obtained about a month before the probe descent, showing
the path the probe took through the atmosphere. The temperature measurements are
from the Atmosphere Structure Instrument and were reported by
\citet{1996Sci...272..844S}. The probe traversed several degrees of longitude
toward the east during the descent, skirting the North Equatorial Belt, a region
where downwelling occurs. The probe measured a very dry atmosphere, with very
little water, which was ultimately attributed to the probe entering a dry ``hot
spot'' of the atmosphere \citep{1998JGR...10322791O,10.2138/rmg.2008.68.10}. This
illustrates the potential problem with an entry probe---a single-point measurement
may not be representative of the entire atmosphere.

\subsection{Radiative Transfer Calculations}

The various types of observations of the atmospheres of the giant planets are often
interpreted using \textit{radiative transfer} modeling. These models calculate the
way light propagates through an atmosphere, subject to solar illumination, deep
thermal radiation, and absorption and scattering by the atmospheric constituents as
well as clouds, aerosols, and hazes. They require inputs such as vertical profiles
of temperature and abundance and the optical properties of the clouds, aerosols,
and hazes.

In general, radiative transfer can be applied in two ways. First, a \textit{forward
model} will take the vertical profiles of abundances, temperature, and any cloud
and aerosol structure to produce a spectrum as it would appear at the top of the
atmosphere. This can be thought of as a prediction of what a particular instrument
would observe. The second way is a \textit{retrieval}, whereby the opposite takes
place, and based on an observed spectrum, the calculations will provide the
composition, temperature, clouds, and aerosol structures that best fit the observed
data. This is a powerful technique that allows constraints on the vertical profiles.

\subsection{Future Prospects}

Both the ESA JUICE mission and the NASA Europa Clipper mission are on their way to
Jupiter, which will provide up-close views of Jupiter's atmosphere in the 2030s.
Ultimately, JUICE will go into orbit around Ganymede, and Europa Clipper will go
into orbit around Europa to explore these enigmatic moons.

The last visit to Uranus and Neptune was by Voyager 2 in 1986 and 1989,
respectively. There is growing consensus in the planetary science community that a
new, Cassini-style flagship mission to either of these planets would provide
significant scientific value, and a Uranus mission was ranked the highest in the
National Academy of Sciences Decadal Strategy for Planetary Science and
Astrobiology 2023--2032 \citep{NAP26522}. In order to perform a gravity assist on
Jupiter, a launch date by 2031 would be required, with an estimated arrival in
2044. To operate a spacecraft over these long durations, far away from the Sun, a
radioisotope thermoelectric generator is required to provide power and heating for
the subsystems. While there is a plutonium shortage, other alternatives are being
explored at ESA using americium (Am). The Uranus Orbiter and Probe mission concept
\citep{simon2021uranus} outlines a payload that includes a descent probe to be
dropped into the atmosphere, providing a ground truth for future measurements, in a
similar way to the Galileo probe on Jupiter.

Lastly, there are a number of 30-m-class telescopes under construction or in the
planning phase. Generally, these offer limited fields of view but provide high
sensitivity and so will be well suited to observing the ice giants, which are often
challenging to observe with smaller telescopes. One example of this new class of
telescope is the Extremely Large Telescope in Chile, which is a 39-m-diameter
telescope that can achieve diffraction-limited spatial resolution.

\section{Conclusion}
\label{sec:conclusion}

The thermal structure, composition, and behavior of the atmospheres of the giant
planets have been outlined in broad strokes. While similarities exist, especially
within the subsets of gas giants and ice giants, significant differences are
present. These differences mean that the four planets exhibit different
atmospheres, and each atmosphere provides a unique laboratory in which to
understand how these systems behave.

Broadly speaking, the atmospheres of the giant planets are convectively driven at
depth by the latent internal heat, and as the atmosphere gets more rarefied and
less opaque with increasing altitudes, radiative processes become more important,
including solar-driven photochemistry in the stratosphere and upper atmosphere.
Instabilities, vortices, and waves are all important for understanding the
small-scale behavior, the generation of convective outbreaks, and the transfer of
energy and momentum within these atmospheres.

A great number of fundamental questions about the atmospheres of the giant planets
remain, and outlined here is a (subjective) list of some of the more pressing ones:

\begin{itemize}
\item What drives the fast zonal winds? How can they remain stable over time? How
(and why) do they decay with altitude? How deep do they penetrate? Why does
Neptune have strong retrograde winds, while Jupiter is dominated by prograde winds?

\item How are the different atmospheric layers coupled on the giant planets? How do
energy and momentum move within and between different layers?

\item Why does Uranus have so little internal heat?

\item What are the abundances of noble gases? They provide important constraints on
the formation process of the Solar System, and so how do these abundances compare
with different formation scenarios?

\item Where do the giant planets sit on the spectrum of giant extrasolar planets?
Are they ``typical'' of this larger sample?
\end{itemize}

\section*{Acknowledgments}

HM was supported by an STFC James Webb Fellowship (ST/W001527/2) at Northumbria
University, UK.

\bibliographystyle{plainnat}
\bibliography{references}

\end{document}